\documentclass{jpsj3}  
\usepackage{txfonts}

\title{Ginzburg-Landau Equations for Coexistent States 
of Superconductivity and Antiferromagnetism in $t-J$ Model
 }

\author{\name{Kazuhiro \surname{KUBOKI}}\thanks{E-mail address: kuboki@kobe-u.ac.jp} 
}
\inst{\address{Department of Physics, Kobe University, 
Kobe 657-8501, Japan} 
}

\abst{
Ginzburg-Landau (GL) equations for the coexistent state of superconductivity 
and antiferromagnetism are derived microscopically from the $t-J$ model 
with extended transfer integrals. 
GL equations and the GL free energy, which are obtained  based on  
the slave-boson mean-field approximation, reflect the electronic structure of the 
microscopic model, especially the evolution of the Fermi surface due to the change 
of the doping rate. 
Thus they are suitable for studying the material dependence 
of the coexistent states in  high-$T_C$ cuprate superconductors.  

 }

\kword{GL theory, t-J model, coexistence, antiferromagnetism}

\begin{document}
\maketitle

\section{Introduction}

The discovery of the coexistence of antiferromagnetism (AF) and 
superconductivity (SC) in multilayer high-$T_c$ cuprates has 
stimulated wide interest. \cite{Kitaoka,Mukuda} 
Antiferromagnetic superexchange interactions in high-$T_C$ cuprate
superconductors, which are strongly correlated electron systems, are thought 
to be the origin of two ordered states; 
thus understanding the condition for coexistence 
may give insight into the mechanism of superconductivity. 

In single-layer and bilayer cuprates such as La- and Y-based compounds,  
it has been well known that AF is easily suppressed by a tiny amount of carrier 
doping.\cite{Keimer,Sanna} 
On the contrary in multilayer systems (in this paper the term "multilayer" will refer 
to three or more layers in a unit cell)  such as 
HgBa$_2$Ca$_4$Cu$_5$O$_{12+y}$, AF survives up to much higher 
doping rate and coexists with SC state. NMR measurements revealed that the 
coexistence was not due to a proximity effect but a genuine phase transition 
within a CuO$_2$ plane.\cite{Kitaoka,Mukuda} 
Multilayer cuprates have flat CuO$_2$ planes with a perfect square lattice 
and are known to be free from disorder in contrast to La- and Y-based compounds. 
Combined with their high $T_C$ of more than 100K,\cite{Iyo} 
multilayer cuprates can be 
viewed as ideal systems to study the mechanism of high $T_C$. 
In this sense it is desirable to explore the nature of the 
coexistent state of AF and SC theoretically. 

Low-energy electronic states of high-$T_C$ cuprates are described by 
the $t-J$ model. \cite{Anderson,Ogata,Lee}
In the case of single-layer and bilayer systems the AF order  
is easily destabilized by strong fluctuations due to low dimensionality.  
Assuming the absence of AF order, 
mean-field  (MF) theories\cite{Kotliar,Suzumura}  based on the slave-boson (SB) 
scheme\cite{Zou} to treat the condition of no double occupancy 
and the gauge theory,\cite{Nagaosa,Lee} which takes into 
account the low-energy fluctuations around mean-fields,  
capture many important properties of single-layer and bilayer high-$T_C$ 
cuprates.   
In multilayer systems, on the other hand, relatively strong three dimensionality 
may stabilize AF order.\cite{KK1,Yamase2}  
This situation can be suitably treated by  MF theories 
for the $t-J$ model by taking AF order into account.  
Actually MF calculations for the $t-J$ model predicted 
that AF survives up to $\delta \lesssim 0.1-0.15$ 
 ($\delta$ being the doping rate) 
and it may coexist with SC,\cite {Inaba,Yamase,KK1,Yamase2} 
and a similar result was obtained by the variational Monte Carlo 
method.\cite{Himeda} 

In this paper, we derive GL equations and the GL free energy 
microscopically from the two-dimensional $t-J$ model with 
extended transfer integrals (extended $t-J$ model) based on the 
SBMF approximation. 
In the MF approach the phase diagram will not be sensitive to the number 
of layers. It is the shape of the Fermi surface, in particular, 
the condition for the nesting that is crucial to determine 
the occurrence of the coexistent state, 
and an electronlike Fermi surface can lead to the experimentally 
observed phase diagram.\cite{KK1,Yamase2} 
In multilayer cuprates we expect that such an electronlike Fermi surface 
may be stabilized as one of the Fermi surface due to strong hybridization 
between layers. 
This is the reason why we treat a single-layer (single-band) model, 
and we simulate the difference of the Fermi surface by including the 
extended transfer integrals. 

The derived GL theory can be used to investigate the spatial dependence 
of the AF and SC order parameters (OPs) in high-$T_C$ cuprates,  
and it may provide information on the electronic states in these systems. 
For example, near the surface or impurity the OPs are suppressed, 
and their recovery to the bulk values will provide the coherence length, 
which reflect the underlying electronic structures of each system. 

Although the GL theory is reliable only qualitatively
except near $T_C$, it can give a simple and intuitive description of
the coexistence and competition of multiple OPs. Thus, it is
complementary to more sophisticated methods such as the
Bogoliubov-de Gennes and quasiclassical Green's function theory. 
Previously 
various  models have been employed to derive GL equations microscopically;   
a continuum\cite{Ren} and tight-binding model\cite{Feder} 
with $s$- and $d$-wave SCOPs, 
Hubbard model with nearest-neighbor attractive interactions,\cite{KKYano} 
a model with a spin generalized BCS term and Heisenberg exchange term,\cite{Dahl}
and the $t-J$ model (without taking AF order into account)\cite{KKSig}. 
The method of deriving GL equations in this work 
is based on that by Gor'kov\cite{Gorkov} 
with the extension to include AF order.\cite{KKYano} 

This paper is organized as follows. In $\S2$, we present the model and treat 
it by the SBMF approximation. GL equations and the GL 
free energy are derived in $\S3$. Section 4 is devoted to summary and discussion.

\section{Model and Mean-Field Approximation}

We consider the extended $t-J$ model on a square lattice whose 
Hamiltonian is given as 
\begin{eqnarray}
\displaystyle H = -\sum_{j,\ell,\sigma} 
t_{j\ell} e^{i\phi_{j\ell}} {\tilde c}^\dagger_{j\sigma} {\tilde c}_{\ell\sigma}
 +J\sum_{\langle j,\ell\rangle} {\bf S}_j\cdot {\bf S}_\ell, 
\end{eqnarray}
where the transfer integrals $t_{j\ell}$ are finite for the first-  ($t$), 
second-  ($t'$), and third-nearest-neighbor bonds ($t''$), and vanish otherwise. 
$J (>0)$ is the antiferromagnetic superexchange interaction
and $\langle j,\ell \rangle$ denotes the nearest-neighbor bonds. 
The magnetic field is taken into account using the Peierls phase 
$\phi_{j,\ell} \equiv \frac{\pi}{\phi_0} \int_j^\ell {\bf A}\cdot d{\bf l}$, 
with ${\bf A}$ and $\phi_0 = \frac{hc}{2e}$ being the vector potential 
and  flux quantum, respectively. 
${\tilde c}_{j\sigma}$ is the electron operator in Fock space without 
double occupancy, and we treat this condition using the SB 
method\cite{Zou}   
by writing ${\tilde c}_{j\sigma}=b_j^\dagger f_{j\sigma}$ under 
the local constraint $\sum_{\sigma}f_{j\,\sigma}^{\dagger}f_{j\,\sigma} 
+ b_j^{\dagger}b_j = 1$ 
at every $j$ site. Here $f_{j\sigma}$ ($b_j$) is a fermion (boson) operator  
that carries spin $\sigma$ (charge $e$); the fermions (bosons) are frequently 
referred to as spinons (holons). 
The spin operator  is expressed as 
$
 {\bf S}_j = \frac{1}{2}\sum_{\alpha,\beta}
f^\dagger_{j\alpha} {\bf \sigma}_{\alpha\beta}f_{j\beta}$. 

We decouple Hamiltonian eq. (1) in the following 
manner. \cite{Inaba,Yamase,KK1,Yamase2}  
The bond order parameters 
$\langle \sum_\sigma f^\dagger_{j\sigma}f_{\ell\sigma} \rangle$ 
and $\langle b^\dagger_j b_\ell\rangle$ 
are introduced and we denote 
$\chi_{j,\ell}/2 =
\langle f^\dagger_{j\uparrow}f_{\ell\uparrow} \rangle 
= \langle f^\dagger_{j\downarrow}f_{\ell\downarrow} \rangle$ 
for the nearest-neighbor bond.
Although the bosons are not condensed in purely two-dimensional systems  
at finite temperature ($T$), 
they are almost condensed at low $T$ and 
for finite carrier doping $\delta (\gtrsim 0.02)$.\cite{Inaba}  
Hence we approximate $\langle b_j \rangle \approx {\sqrt \delta}$ and 
$\langle b^\dagger_ib_j\rangle \approx  \delta$. 
The magnetization is defined by 
$m_j = \frac{1}{2}\langle f^\dagger_{j\uparrow}f_{j\uparrow}
- f^\dagger_{j\downarrow}f_{j\downarrow} \rangle $,  
and the superconducting  OP  on the bond $\langle j,\ell\rangle$ 
 (under the assumption of the 
Bose condensation of holons) is given as 
$\Delta_{j,\ell} = \langle f_{j\uparrow}f_{\ell\downarrow}\rangle$. 

Fluctuations around the mean-field solutions in the slave-boson scheme 
can be treated as the gauge field. It is known that this gauge field 
may affect the physical properties of the solutions in a serious way.\cite{Nagaosa}  
However, in the SC and AF states the effect of the gauge field is  
strongly suppressed.\cite{LeeNag,DKim}
Since we are interested in these ordered states, 
we do not consider the effect of gauge-field fluctuations. 

In the following we are mainly interested in a region around the tetracritical point 
where the four states, AF, $d_{x^2-y^2}$-wave SC, their coexistence, 
and the normal states become identical. 
The onset temperature of the bond OPs is much higher than 
that for AF ($T_N$) and SC ($T_C$) in this doping region, 
so that they are almost independent of temperature near the tetracritical point. 
We consider only the spatial variations of $m_j$ and  $\Delta_{j,\ell}$ 
assuming that $\chi_{j,\ell}$ is uniform in space. 
(Hereafter we denote it  as $\chi$.) 
Then the mean-field Hamiltonian is given as 
\begin{equation}\begin{array}{rl}
 H_{MFA} = & \displaystyle -\sum_{j,\sigma}\Big[
\sum_{\delta=\pm x,\pm y} \big(t\delta e^{i\phi_{j+\delta,j}} 
+ \frac{3J}{8}\chi\big)  f^\dagger_{j+\delta,\sigma}f_{j\sigma} 
+ t'\delta\sum_{\delta=\pm x \pm y} 
e^{i\phi_{j+\delta,j}} f^\dagger_{j+\delta,\sigma}f_{j\sigma} \\  
+ & \displaystyle t''\delta\sum_{\delta=\pm 2x, \pm 2y}
 e^{i\phi_{j+\delta,j}}  f^\dagger_{j+\delta,\sigma}f_{j\sigma}  \Big] 
 -  \mu\sum_{j,\sigma} f^\dagger_{j\sigma}f_{j\sigma}
+\frac{J}{2}\sum_j\sum_{\delta=\pm x,\pm y} 
m_{j+\delta}  \big(f^\dagger_{j\uparrow}f_{l\uparrow} 
- f^\dagger_{j\downarrow}f_{l\downarrow}\big)  \\
+ & \displaystyle  \frac{J}{2}\sum_j \sum_{\delta=\pm x,\pm y}
\big[\Delta_{j,j+\delta}\big(f^\dagger_{j+\delta\uparrow}f^\dagger_{j\downarrow} 
-\frac{1}{2}f^\dagger_{j+\delta\downarrow}f^\dagger_{j\uparrow}\big) + h.c.\big] + E_0,  
\end{array}\end{equation}
with 
\begin{equation}
E_0= - J \sum_{\langle j,\ell \rangle} m_j m_\ell
+J\sum_{\langle j,\ell\rangle}\Big(\frac{1}{2}\Delta_{j,\ell}\Delta^*_{\ell,j}
+\frac{1}{4}|\Delta_{j,\ell}|^2\Big).
\end{equation}

First we solve the self-consistency equations for $\chi$ and the 
chemical potential $\mu$  in the absence of $m$, $\Delta$, and ${\bf A}$. 
Self-consistency equations that determine $\chi$ and $\mu$ as functions of 
$T$ and $\delta$  are given as 
\begin{equation}
\displaystyle \chi =  \frac{1}{N}\sum_p (\cos p_x+\cos p_y)f(\xi_p) , \ \ 
\delta=  1 - \frac{2}{N}\sum_pf(\xi_p), 
\end{equation}
where $\xi_p = -(2t\delta + \frac{3J}{4}\chi)(\cos p_x+\cos p_y) 
 -4t'\delta\cos p_x \cos p_y - 2t''\delta(\cos 2p_x+\cos 2p_y) -\mu$,  
with $f$ and $N$ being the Fermi function and 
the total number of lattice sites, respectively.
(Lattice constant is taken to be unity.)
In the next section we will carry out the GL expansion to obtain GL 
equations for $m$ and $\Delta$. 

For the values of $t'$ and $t''$  which reproduce the experimentally 
obtained phase diagram, incommensurate (IC) as well as commensurate 
(C) AF order may be possible around the tetracritical point depending on the 
choice of the parameters.\cite{Yamase2} 
(There are several distinct parameter sets which lead to similar phase diagrams.)
Experimentally, since the NMR does not directly discriminate different ordering 
patterns of magnetism, at present it is not clear whether ICAF order exists.  
Then we will consider only the CAF state as a feasible candidate.

\section{\label{sec:GL}GL Equations and GL Free Energy}
In this section we derive GL equations and the GL free energy. 
The procedure is essentially the same as that used in ref.20. 
Coupled equations for Green's functions 
$G_\sigma(j,\ell,\tau) =  
-\langle T_\tau f_{j\sigma}(\tau)f_{\ell\sigma}^\dagger\rangle$
 and $F^\dagger_{\sigma\sigma'}(j,\ell,\tau) = 
-\langle T_\tau f_{j\sigma}^\dagger(\tau)f_{\ell\sigma'}^\dagger\rangle$ 
can be derived from their equations of motion (Gor'kov equations) as 

\begin{equation}\begin{array}{rl}
 G_\uparrow(j,\ell,i\varepsilon_n) = & \displaystyle {\tilde G}_0(j,\ell,i\varepsilon_n)  
 + \frac{J}{2}\sum_{k,\delta_1} {\tilde G}_0(j,k,i\varepsilon_n)
 \\ & \displaystyle \times
 \Big[\Big(\Delta_{k+\delta_1,k}+\frac{1}{2}\Delta_{k,k+\delta_1}\Big) 
F^\dagger_{\downarrow\uparrow}(k+\delta_1,\ell,i\varepsilon_n) 
 +m_{k+\delta_1} 
G_\uparrow(k,\ell,i\varepsilon_n)\Big], \\ 
G_\downarrow(j,\ell,i\varepsilon_n) = 
& \displaystyle {\tilde G}_0(j,\ell,i\varepsilon_n) 
- \frac{J}{2}\sum_{k,\delta_1}{\tilde G}_0(j,k,i\varepsilon_n) 
 \\ & \displaystyle \times
\Big[\Big(\Delta_{k,k+\delta_1}+\frac{1}{2}\Delta_{k+\delta_1,k}\Big) 
F^\dagger_{\uparrow\downarrow}(k+\delta_1,\ell,i\varepsilon_n) 
+ m_{k+\delta_1} 
G_\downarrow(k,\ell,i\varepsilon_n)\Big], \\ 
F^\dagger_{\downarrow\uparrow}(j,\ell,i\varepsilon_n) = & \displaystyle
- \frac{J}{2}\sum_{k,\delta_1} {\tilde G}_0(k,j,-i\varepsilon_n) 
 \\ & \displaystyle \times 
\Big[\Big(\Delta^*_{k,k+\delta_1}+\frac{1}{2}\Delta^*_{k+\delta_1,k}\Big) 
G_\uparrow(k+\delta_1,\ell,i\varepsilon_n) 
+ m_{k+\delta_1} 
F^\dagger_{\downarrow\uparrow}(k,\ell,i\varepsilon_n)\Big],  \\ 
F^\dagger_{\uparrow\downarrow}(j,\ell,i\varepsilon_n) = & \displaystyle
\frac{J}{2}\sum_{k,\delta_1} {\tilde G}_0(k,j,-i\varepsilon_n) 
\\ & \displaystyle \times 
\Big[\Big(\Delta^*_{k+\delta_1,k}+\frac{1}{2}\Delta^*_{k,k+\delta_1}\Big) 
G_\downarrow(k+\delta_1,\ell,i\varepsilon_n) 
+ m_{k+\delta_1} 
F^\dagger_{\uparrow\downarrow}(k,\ell,i\varepsilon_n)\Big],  \\ 
\end{array}\end{equation} 
where the summation on $\delta_1$ ($k$) is over 
 $\pm {\hat x}$ and $\pm {\hat y}$ (all sites). 
Here,  ${\tilde G}_0(j,\ell,i\omega_n)$  is  Green's function for
the system without $\Delta$ and $m$  but with ${\bf A}$. 
${\tilde G}_0(j,\ell,i\omega_n)$ is related to  Green's function for the 
system without ${\bf A}$, $G_0$, as 
$ {\tilde G}_0(j,\ell,i\varepsilon_n) 
\sim   G_0(j,\ell,i\varepsilon_n) e^{i\phi_{j,\ell}}$, with  
$G_0(j,\ell,i\varepsilon_n)$ being the Fourier transform of 
$G_0({\bf p},i\varepsilon_n) = 1/(i\varepsilon_n-\xi_p)$.  
In the expression of $\xi_p$,  
the bond order parameter $\chi$ and the chemical potential $\mu$ 
determined by eq.(4) are substituted.

Spin-singlet and spin-triplet SCOPs on the bond $(j,j+\eta)$ are expressed 
in terms of  Green's functions $F_{\uparrow\downarrow}^\dagger$ 
and $F_{\downarrow\uparrow}^\dagger$, 
\begin{equation}\begin{array}{rl}
\displaystyle (\Delta_\eta^{(S)}(j))^* \equiv   & \displaystyle \frac{1}{2} \langle 
f_{j\uparrow}f_{j+\eta\downarrow} - f_{j\downarrow}f_{j+\eta\uparrow}\rangle^*  
=   \frac{1}{2} \big(\Delta_{j,j+\eta}+\Delta_{j+\eta,j}\big)^* \\
= & \displaystyle \frac{T}{2} \sum_{\varepsilon_n}
\Big[F^\dagger_{\uparrow\downarrow}(j+\eta,j,i\varepsilon_n) 
- F^\dagger_{\downarrow\uparrow}(j+\eta,j,i\varepsilon_n)\Big], \\
\displaystyle  (\Delta_\eta^{(T)}(j))^* \equiv & \displaystyle\frac{1}{2} \langle 
f_{j\uparrow}f_{j+\eta\downarrow} + f_{j\downarrow}f_{j+\eta\uparrow}\rangle^*
=   \frac{1}{2} \big(\Delta_{j,j+\eta}-\Delta_{j+\eta,j} \big)^* \\
= &  \displaystyle  -\frac{T}{2} \sum_{\varepsilon_n}
\Big[F^\dagger_{\uparrow\downarrow}(j+\eta,j,i\varepsilon_n) 
+ F^\dagger_{\downarrow\uparrow}(j+\eta,j,i\varepsilon_n)\Big], \\ 
\end{array}\end{equation}
and the staggered magnetization $M_j  \equiv m_j e^{i{\bf Q}\cdot{\bf r}_j}$ 
(${\bf Q} = (\pi,\pi)$)
is similarly given using  $G_{\uparrow}$ 
and $G_{\downarrow}$, 
\begin{equation}\begin{array}{rl}
 M_j \equiv & \displaystyle 
 \frac{1 }{2}  \langle f^\dagger_{j\uparrow}f_{j\uparrow} 
- f^\dagger_{j\downarrow}f_{j\downarrow} \rangle   e^{i{\vec Q}\cdot{\vec r}_j}  \\
= & \displaystyle \frac{T}{2} \sum_{\varepsilon_n}
\big[G_\uparrow(j,j,i\varepsilon_n)-G_\downarrow(j,j,i\varepsilon_n) \big]
 e^{i{\vec Q}\cdot{\vec r}_j}. 
\end{array}\end{equation}
We substitute eq. (5) into eqs. (6) and (7) iteratively 
and keep the terms up to the third order in OPs.  
In the coexistent state of  AF and SC, spin-triplet SCOPs that oscillate in a similar 
manner as the staggered magnetization may 
occur\cite{Fenton,Mura1,Mura2,Kyung,Apens}, 
and they are called the $\pi$-triplet SCOPs.  
The SCOPs of each symmetry, $\Delta_s$ ($s$-wave), $\Delta_d$ ($d$-wave), 
and $\Delta^{(\pi T)}_{px(y)}$ ($\pi$-triplet  $px(y)$-wave),  
 can be constructed by making a linear combination of eq.(6), 
\begin{equation}\begin{array}{rl}
 \Delta_s(j) = &  \displaystyle \frac{1}{4} \sum_{\eta=\pm{\hat x},\pm{\hat y}}
 \Delta_\eta^{(S)}(j) ,  \ \ 
 \Delta_d(j) =  \frac{1}{4} \Big[\sum_{\eta=\pm{\hat x}}  \Delta_\eta^{(S)}(j) 
 - \sum_{\eta=\pm{\hat y}}  \Delta_\eta^{(S)}(j)\Big],  
 \\
 & \displaystyle \Delta^{(\pi T)}_{px(y)}(j) = 
\frac{1}{2} \big[\Delta^{(\pi T)}_{{\hat x}(y)}(j) + 
\Delta^{(\pi T)}_{-{\hat x}(y)}(j)\big].  
\end{array}\end{equation}

Assuming that the SCOPs and  $M$ are slowly varying,  
we take a continuum limit.  The OPs in the linear terms are 
expanded in powers of derivatives up tp the second order, 
and the Peierls phase is also expanded 
in powers of ${\bf  A}$ to the same order.  
Then after straightforward but lengthy calculations we get the 
following GL equations:  

\begin{equation}\begin{array}{rl}
& \displaystyle 
\alpha_s \Delta_s + 2\beta_s |\Delta_s|^2\Delta_s 
- K_s (D_x^2+D_y^2) \Delta_s - K_{ds}(D_x^2-D_y^2)\Delta_d \\
+  & \displaystyle \gamma_1|\Delta_d|^2\Delta_s + 2\gamma_2\Delta_d^2\Delta_s^*
+ \gamma_3(|\Delta^{(\pi T)}_{px}|^2+|\Delta^{(\pi T)}_{py}|^2)\Delta_s 
+ 2\gamma_5((\Delta^{(\pi T)}_{px})^2+(\Delta^{(\pi T)}_{py})^2)\Delta_s^* \\
+  & \displaystyle \gamma_7(|\Delta^{(\pi T)}_{px}|^2
-|\Delta^{(\pi T)}_{py}|^2)\Delta_d
+ \gamma_8((\Delta^{(\pi T)}_{px})^2-(\Delta^{(\pi T)}_{py})^2)\Delta_d^* 
+ \gamma_9(\Delta^{(\pi T)*}_{px}\Delta^{(\pi T)}_{py} + c.c.)\Delta_s \\
+ & \displaystyle 2\gamma_{11}\Delta^{(\pi T)}_{px}\Delta^{(\pi T)}_{py}\Delta_s^*
+ \gamma_{ms}M^2\Delta_s + \gamma_{spm}M
(\Delta^{(\pi T)}_{px}+\Delta^{(\pi T)}_{py}) 
= 0, 
\end{array}\end{equation}
\begin{equation}\begin{array}{rl}
& \displaystyle 
\alpha_d \Delta_d + 2\beta_d |\Delta_d|^2\Delta_d 
- K_d (D_x^2+D_y^2) \Delta_d - K_{ds}(D_x^2-D_y^2)\Delta_s \\
+  & \displaystyle \gamma_1|\Delta_s|^2\Delta_d + 2\gamma_2\Delta_s^2\Delta_d^*
+ \gamma_4(|\Delta^{(\pi T)}_{px}|^2+|\Delta^{(\pi T)}_{py}|^2)\Delta_d 
+ 2\gamma_6((\Delta^{(\pi T)}_{px})^2+(\Delta^{(\pi T)}_{py})^2)\Delta_d^* \\
+  & \displaystyle \gamma_7(|\Delta^{(\pi T)}_{px}|^2
-|\Delta^{(\pi T)}_{py}|^2)\Delta_s 
+ \gamma_8((\Delta^{(\pi T)}_{px})^2-(\Delta^{(\pi T)}_{py})^2)\Delta_s^* 
+ \gamma_{10}(\Delta^{(\pi T)*}_{px}\Delta^{(\pi T)}_{py} + c.c.)\Delta_d \\
+ & \displaystyle 2\gamma_{12}\Delta^{(\pi T)}_{px}\Delta^{(\pi T)}_{py}\Delta_d^*
+ \gamma_{md}M^2\Delta_d 
+ \gamma_{dpm}M(\Delta^{(\pi T)}_{px}-\Delta^{(\pi T)}_{py}) 
= 0, 
\end{array}\end{equation}
\begin{equation}\begin{array}{rl}
& \displaystyle 
\alpha_{p1} \Delta^{(\pi T)}_{px(y)} + \alpha_{p2} \Delta^{(\pi T)}_{py(x)} 
+ 2\beta_p |\Delta^{(\pi T)}_{px(y)}|^2\Delta^{(\pi T)}_{px(y)} \\
- & \displaystyle K_{p1}D_{x(y)}^2\Delta^{(\pi T)}_{px(y)}
-K_{p2}D_{y(x)}^2\Delta^{(\pi T)}_{px(y)} 
- K_{p3}(D_x^2+D_y^2)\Delta^{(\pi T)}_{py(x)}  \\ 
+ & \displaystyle  \gamma_{p1}|\Delta^{(\pi T)}_{py(x)}|^2\Delta^{(\pi T)}_{px(y)} 
+ 2\gamma_{p2}(\Delta^{(\pi T)}_{py(x)})^2\Delta^{(\pi T)*}_{px(y)}  \\
+ & \displaystyle \gamma_{p3}(2|\Delta^{(\pi T)}_{px(y)}|^2\Delta^{(\pi T)}_{py(x)}
+ (\Delta^{(\pi T)}_{px(y)})^2\Delta^{(\pi T)*}_{py(x)}
+|\Delta^{(\pi T)}_{py(x)}|^2\Delta^{(\pi T)}_{py(x)}) \\
+ & \displaystyle (\gamma_3|\Delta_s|^2+\gamma_4|\Delta_d|^2) 
\Delta^{(\pi T)}_{px(y)}
+ 2(\gamma_5\Delta_s^2+\gamma_6\Delta_d^2) \Delta^{(\pi T)*}_{px(y)} \\
\pm & \displaystyle \gamma_7(\Delta_s\Delta_d^*+c.c.)\Delta^{(\pi T)}_{px(y)} 
\pm 2\gamma_8 \Delta_s\Delta_d\Delta^{(\pi T)*}_{px(y)} 
+(\gamma_9|\Delta_s|^2+\gamma_{10}|\Delta_d|^2)\Delta^{(\pi T)}_{py(x)} 
\\
+ & \displaystyle (\gamma_{11}\Delta_s^2+\gamma_{12}\Delta_d^2)
\Delta^{(\pi T)*}_{py(x)}
+(\gamma_{mp1}\Delta^{(\pi T)}_{px(y)}
+\gamma_{mp2}\Delta^{(\pi T)}_{py(x)})M^2 \\
+ & \displaystyle (\gamma_{spm}\Delta_s \pm \gamma_{dpm}\Delta_d)M
= 0,  
\end{array}\end{equation}
\begin{equation}\begin{array}{rl}
& \displaystyle 
\alpha_m M+ 2\beta_m M^3 -K_m(\nabla_x^2+\nabla_y^2)M \\
+ & \displaystyle  (\gamma_{ms}|\Delta_s|^2+\gamma_{md}|\Delta_d|^2)M
+ [\gamma_{mp1}(|\Delta^{(\pi T)}_{px}|^2+|\Delta^{(\pi T)}_{py}|^2)
+\gamma_{mp2}(\Delta^{(\pi T)}_{px}\Delta^{(\pi T)*}_{py}+c.c.)]M \\
+ & \displaystyle \frac{1}{2}  
\gamma_{spm}[\Delta_s^*(\Delta^{(\pi T)}_{px}+\Delta^{(\pi T)}_{py}) +c.c.]
 +  \frac{1}{2} \gamma_{dpm}[\Delta_d^*(\Delta^{(\pi T)}_{px}
 -\Delta^{(\pi T)}_{py})+c.c.]  
 = 0,
\end{array}\end{equation}
where the coefficients appearing in eqs. (9)-(12) are 
given in the Appendix, and  ${\bf D}$ is the gauge-invariant gradient 
defined as $ {\bf D} \equiv {\bf \nabla} +\frac{2\pi i}{\phi_0}{\bf A}$.  
Equations (9)-(12) are the coupled equations that determine SCOPs and the 
staggered magnetization self-consistently. 

The GL free energy $F$ up to the fourth order in OPs 
can be obtained from the above GL equations in such a way that the 
variations of $F$ with respect to OPs reproduce eqs. (9)-(12).  
The results are written as follows: 
\begin{equation}\begin{array}{rl}
 \displaystyle F = & \displaystyle F_S  + F_T + F_{ST} +F_M + F_{SM}
 + F_{TM} + F_{STM}, \\
 F_S = &\displaystyle  \int d^2{\bf r} \Big[
 \alpha_s |\Delta_s|^2 + \beta_s |\Delta_s|^4  + K_s |{\vec D} \Delta_s|^2 
 + \alpha_d |\Delta_d|^2+ \beta_d |\Delta_d|^4 +  K_d |{\vec D} \Delta_d|^2 \\
 & \displaystyle + \gamma_1 |\Delta_s|^2|\Delta_d|^2 
 + \gamma_2 \big(\Delta_d^2(\Delta_s^*)^2 + c.c.\big)  \\
& + K_{ds} \big((D_x\Delta_d)(D_x\Delta_s)^{*} 
 - (D_y\Delta_d)(D_y\Delta_s)^{*} + c.c. \big)\Big],   \\
 F_T =   &\displaystyle \int d^2{\bf r} \Big[
\alpha_{p1}\big(|\Delta_{px}^{(\pi T)}|^2 + |\Delta_{py}^{(\pi T)}|^2\big) 
+ \alpha_{p2}\big(\Delta_{px}^{(\pi T)}(\Delta_{py}^{(\pi T)})^* + c.c\big) 
+ \beta_p\big(|\Delta_{px}|^4 + |\Delta_{py}|^4\big)  \\
& \displaystyle + \gamma_{p1}
|\Delta_{px}^{(\pi T)}|^2|\Delta_{py}^{(\pi T)}|^2  
+  \gamma_{p2}\big((\Delta_{px}^{(\pi T)})^2(\Delta_{py}^{(\pi T)*})^2 
+ c.c.\big) \\
& \displaystyle 
+  \gamma_{p3}
\big(|\Delta_{px}^{(\pi T)}|^2 + |\Delta_{py}^{(\pi T)}|^2\big) 
\big(\Delta_{px}^{(\pi T)}(\Delta_{py}^{(\pi T)})^*+ c.c.\big) \\
& \displaystyle +  K_{p1}\big(|D_x\Delta_{px}^{(\pi T)}|^2 
+ |D_y\Delta_{py}^{(\pi T)}|^2\big)
+  K_{p2}\big(|D_y\Delta_{px}^{(\pi T)}|^2 
+ |D_x\Delta_{py}^{(\pi T)}|^2\big) \\
& \displaystyle + K_{p3}\big((D_x\Delta_{px}^{(\pi T)})^{*}
(D_x\Delta_{py}^{(\pi T)})
+ (D_y\Delta_{px}^{(\pi T)})^{*}(D_y\Delta_{py}^{(\pi T)}) + c.c.\big)\Big], \\
 F_{ST} = &\displaystyle  \int d^2{\bf r} \Big[
 \big(|\Delta_{px}^{(\pi T)}|^2 + |\Delta_{py}^{(\pi T)}|^2\big)
(\gamma_3|\Delta_s|^2 + \gamma_4|\Delta_d|^2) \\
& \displaystyle +  
\big\{\big((\Delta_{px}^{(\pi T)})^2 + (\Delta_{py}^{(\pi T)})^2\big)
(\gamma_5(\Delta_s^{*})^2 + \gamma_6(\Delta_d^{*})^2) + c.c \big\}
\\ 
& \displaystyle +  \gamma_7 
\big(|\Delta_{px}^{(\pi T)}|^2 - |\Delta_{py}^{(\pi T)}|^2\big)
\big(\Delta_s^{*}\Delta_d + c.c.\big) 
+  \gamma_8 
\big\{\big((\Delta_{px}^{(\pi T)})^2 - (\Delta_{py}^{(\pi T)})^2\big)
\Delta_s^{*}\Delta_d^{*} + c.c.\big\}\Big] \\
& \displaystyle +  
\big((\Delta_{px}^{(\pi T)})^*\Delta_{py}^{(\pi T)}+c.c.\big)
(\gamma_9 |\Delta_s|^2 + \gamma_{10} |\Delta_d|^2) \\  
& \displaystyle +  
\big\{\Delta_{px}^{(\pi T)}\Delta_{py}^{(\pi T)}
(\gamma_{11}(\Delta_s^*)^2 +  \gamma_{12}(\Delta_d^*)^2)
+c.c. \big\}\Big], \\
 F_M =  &\displaystyle \int d^2{\bf r} \Big[
 \alpha_m  M^2 +  \beta_m M^4 
+  K_m \big(\nabla M\big)^2\big], \\ 
 F_{SM}  = &\displaystyle \int d^2{\bf r} \Big(
 \gamma_{ms} M^2 |\Delta_s|^2 
+ \gamma_{md} M^2 |\Delta_d|^2 \Big), \\
F_{TM}  = &\displaystyle \int d^2{\bf r} \Big[ 
 \gamma_{mp1} M^2\Big(|\Delta_{px}^{(\pi T)}|^2  
+ |\Delta_{py}^{(\pi T)}|^2\Big) 
 \\ & \displaystyle 
+  \gamma_{mp2} M^2 \Big(\Delta_{px}^{(\pi T)}
(\Delta_{py}^{(\pi T)})^* + c.c.\Big)\Big],  \\
\displaystyle F_{STM}  =  &\displaystyle \int d^2{\bf  r}  \Big[
 \gamma_{spm} M \Delta_s\big(\Delta_{px}^{(\pi T)} 
+ \Delta_{py}^{(\pi T)}\big)^*  
\\ & \displaystyle 
+  \gamma_{dpm} M \Delta_d\big(\Delta_{px}^{(\pi T)} 
- \Delta_{py}^{(\pi T)}\big)^* 
+ c.c.\Big]. 
\end{array}\end{equation}
Here, $F_S$, $F_T$, and $F_M$ are the free energy for the singlet and 
$\pi$-triplet SCOPs,  and 
the staggered magnetization, respectively, while $F_{ST}$, $F_{SM}$, $F_{TM}$, and 
$F_{STM}$ describe their couplings. 
Note that $F$ is invariant under all the symmetry operations of the 
square lattice. 
$F_{SM}$ and $F_{TM}$ are the 
usual terms to represent the competition of SCOPs and $M$. 
$F_{STM}$ is a cubic term that couples spin-singlet SCOPs, staggered 
magnetization, and $\pi$-triplet  SCOPS, and it induces $\pi$-triplet  
SCOPs in the coexistent state of AF and SC. 
Generally in the coexistent state of ferromagnetism and spin-singlet SC state, 
spin-triplet SCOPs may occur when OPs are not uniform in 
space.\cite{KK2,Berg2,Esch,Buz,Berg} 
In the GL theory this can be explained by a cubic term 
that has a gradient coupling of spin-singlet, triplet SCOPs, and the 
magnetization $m$.\cite{KKYano} 
In the AF state magnetization $m$ is oscillating 
(though the staggered magnetization $M$ is uniform) even in a uniform case, 
and thus $\pi-$triplet SCOP can arise irrespective of the spatial dependence of 
OPs. 

The important point of the present results 
is that the coefficients appearing in GL equations and the
GL free energy are determined microscopically. 
These values depend on the parameters of the microscopic model 
and they reflect the evolution of the shape of the Fermi surface.  
This property can be used to study the material dependence of 
the coexistent states in various multilayer high-$T_C$ cuprates.

\section{\label{sec:summary}Summary and Discussion}

We have derived GL equations and the GL free energy microscopically 
from the extended $t-J$ model using the slave-boson mean-field 
approximation. 
The derived GL theory can be used to investigate the spatial dependence 
of the AF and SC order parameters in high-$T_C$ cuprate  
superconductors. 
By analyzing the spatial variations of order parameters using the present results, 
information on the electronic states of high-$T_C$ cuprates 
may be extracted. 

A typical example to be studied is the state near the surface or impurity. 
The interface states of heterostructures composed of cuprate 
superconductors and magnetic materials are also worth studying.
There the coexistence and competition of superconductivity 
and magnetism can occur in various ways depending on the materials used. 

Numerical study of the GL equations for the above situations 
assuming various band structure (by choosing the extended transfer integrals) 
may be interesting, and this problem will be examined separately. 


\begin{acknowledgment}
The author thanks H. Yamase for useful discussions. 

\end{acknowledgment}

\appendix
\section{Coefficients in GL Equations and GL Free Energy}
The coefficients appearing in GL equations [eqs.(9)-(12)] 
and the GL free energy [eq.(13)] are given as follows:  
\begin{equation}\begin{array}{rl}
& \displaystyle\alpha_{s(d)} = 3J \Big(1-\frac{3J}{4N}\sum_p 
I_1(p) \omega_{s(d)}^2 \Big), \\
& \displaystyle  \beta_{s(d)} = \frac{81J^4}{32N}\sum_p 
I_2(p) \omega_{s(d)}^4, \\
& \displaystyle  \gamma_1= \frac{81J^4}{8N}\sum_p 
I_2(p) \omega_s^2 \omega_d^2, 
\ \ \  \gamma_2 = \frac{1}{4} \gamma_1, \\
& \displaystyle  K_{s(d)}=  \frac{9J^2}{8N} \sum_p I_2(p) 
\Big(\frac{\partial \xi_p}{\partial p_x}\Big)^2 \omega_{s(d)}^2,  \\
& \displaystyle  K_{ds}= \frac{9J^2}{8N} \sum_p I_2(p) 
\Big(\frac{\partial \xi_p}{\partial p_x}\Big)^2 \omega_s \omega_d, \\
& \displaystyle  \alpha_{p1} = -\frac{J}{2} \Big(1+\frac{J}{2N}\sum_p 
I_3(p) \cos^2p_x \Big) , \\
& \displaystyle
\alpha_{p2} = -\frac{J^2}{4N}\sum_p 
I_3(p) \cos p_x \cos p_y,  \\
& \displaystyle  \beta_p = \frac{J^4}{32N}\sum_p 
I_4(p) \cos^4p_x,\\
& \displaystyle  \gamma_{p1} = \frac{J^4}{8N}\sum_p
I_4(p) \cos^2p_x\cos^2p_y, 
\ \ \  \gamma_{p2} = \frac{1}{4} \gamma_{p1}, \\
& \displaystyle  \gamma_{p3} = \frac{J^4}{16N}\sum_p
I_4(p) \cos^3p_x\cos p_y, \\
& \displaystyle  K_{p1(2)} = - \frac{J^2}{8N}\sum_p
I_4(p) \Big(\frac{\partial \xi_p}{\partial p_x}\Big)^2 \cos^2p_{x(y)}, \\ 
 & \displaystyle 
K_{p3} = - \frac{J^2}{8N}\sum_p
I_4(p) \Big(\frac{\partial \xi_p}{\partial p_x}\Big)^2 \cos p_x\cos p_y, \\
& \displaystyle  \gamma_{3(4)} = \frac{9J^4}{8N} \sum_p I_5(p) 
\omega_{s(d)}^2 \cos^2p_x, \\
& \displaystyle  \gamma_{5(6)} = \frac{9J^4}{32N} \sum_p I_6(p) 
\omega_{s(d)}^2 \cos^2p_x, \\
& \displaystyle \gamma_7 = \frac{9J^4}{8N} \sum_p I_5(p) 
\omega_s\omega_d \cos^2p_x, \\
& \displaystyle \gamma_8 = \frac{9J^4}{16N} \sum_p I_6(p) 
\omega_s\omega_d \cos^2p_x, \\
& \displaystyle  \gamma_{9(10)} = \frac{9J^4}{8N} \sum_p I_5(p) 
\omega_{s(d)}^2 \cos p_x \cos p_y, \\
& \displaystyle  \gamma_{11(12)} = \frac{9J^4}{16N} \sum_p I_6(p) 
\omega_{s(d)}^2 \cos p_x \cos p_y, \\
&  \displaystyle  \alpha_m = 2J\Big(1+\frac{2J}{N}\sum_p I_7(p)\Big), \\
              \end{array}\end{equation}
             \begin{equation}\begin{array}{rl}

&  \displaystyle  \beta_m = \frac{8J^4}{N}\sum_p I_8(p), \\

&  \displaystyle   K_m = \frac{4J^2}{N}\sum_p I_8(p) 
\Big(\frac{\partial \xi_p}{\partial p_x}\Big)^2,  \\

&  \displaystyle  \{ \gamma_{ms}, \  \gamma_{md}\} 
 = - \frac{9J^4}{N} \sum_p [2I_9(p)+I_6(p)]
\{ \omega_s^2, \omega_d^2\}, \\

&  \displaystyle  \{\gamma_{mp1}, \  \gamma_{mp2}\}                
 = - \frac{J^4}{N} \sum_p [2I_{10}(p)+I_6(p)]    
\{\cos^2p_x, \cos p_x\cos p_y\}, \\
&   \displaystyle  \{ \gamma_{spm}, \ \gamma_{dpm}\} 
 = -\frac{3J^3}{N} \sum_p I_{11}(p) \cos p_x 
\{\omega_s,\omega_d\},  
\end{array}\end{equation}
where $\omega_s=\cos p_x+\cos p_y$ and $\omega_d=\cos p_x -\cos p_y$, 
and the summation on $p$ is taken over the first Brillouin zone. 
 The functions appearing in the integrands are defined as 
\begin{equation}\begin{array}{rl}
I_1(p) = & \displaystyle 
T\sum_{\varepsilon_n} G_0(p,i\varepsilon_n)G_0(p,-i\varepsilon_n), \\
I_2(p) = & \displaystyle 
T\sum_{\varepsilon_n} G_0^2(p,i\varepsilon_n)G_0^2(p,-i\varepsilon_n), \\
I_3(p)  = & \displaystyle 
T\sum_{\epsilon_n}G_0(p,-i\varepsilon_n)G_0(p+Q,i\varepsilon_n),  \\
%
I_4(p) = & \displaystyle 
T\sum_{\epsilon_n}G_0^2(p,i\varepsilon_n)G_0^2(p+Q,-i\varepsilon_n), \\
%
I_5(p) = & \displaystyle T\sum_{\epsilon_n}
G_0^2(p,i\varepsilon_n)G_0(p,-i\varepsilon_n)G_0(p+Q,-i\varepsilon_n), \\
%
I_6(p) = & \displaystyle 
T\sum_{\epsilon_n}G_0(p,i\varepsilon_n)G_0(p,-i\varepsilon_n)
G_0(p+Q,i\varepsilon_n)G_0(p+Q,-i\varepsilon_n), \\
%
I_7(p) = & \displaystyle 
T\sum_{\epsilon_n}G_0(p,i\varepsilon_n)G_0(p+Q,i\varepsilon_n), \\
%
I_8(p) = & \displaystyle 
T\sum_{\epsilon_n}G_0^2(p,i\varepsilon_n)G_0^2(p+Q,i\varepsilon_n), \\
%
I_9(p) = & \displaystyle T\sum_{\epsilon_n}
G_0^2(p,i\varepsilon_n)G_0(p,-i\varepsilon_n)G_0(p+Q,i\varepsilon_n), \\
%
I_{10}(p) = & \displaystyle T\sum_{\epsilon_n}
G_0^2(p,i\varepsilon_n)G_0(p+Q,i\varepsilon_n)G_0(p+Q,-i\varepsilon_n),  \\
%
I_{11}(p) = & \displaystyle T\sum_{\epsilon_n}
G_0(p,i\varepsilon_n)G_0(p,-i\varepsilon_n)G_0(p+Q,i\varepsilon_n). \\
\end{array}\end{equation}



\begin{thebibliography}{9}

\bibitem{Kitaoka} Y. Kitaoka,  H. Mukuda, S. Shimizu, S. Tabata, 
P. M. Shirage, and A. Iyo:     
J. Phys. Chem. Solids {\bf 72} (2011) 486. 

\bibitem{Mukuda} H. Mukuda, S. Shimizu, A. Iyo,    and 
Y. Kitaoka: J. Phys. Soc. Jpn. {\bf 81} (2012) 011008.  

\bibitem{Keimer} B. Keimer, N. Belk, B. J. Birgeneau, A. Cassanho, M. Greven, 
M. A. Kastner, A. Aharony, Y. Endoh, R. W. Erwin, and G. Shirane: 
Phys. Rev. B{\bf 46} (1992) 14034. 

\bibitem{Sanna} S. Sanna, G. Allodi, G. Concas, A. D. Hillier, and R. De Renzi: 
Phys. Rev. Lett. {\bf 93} (2004) 207001. 

\bibitem{Iyo} A, Iyo, Y. Tanaka, H. Kito, Y. Kodama, P. M. Shirage, D. D. Shivagan, 
H. Matsuhata, K. Tokiwa, and T. Watanabe: 
J. Phys. Soc. Jpn. {\bf 76} (2007) 094711.  

\bibitem{Anderson} P. W. Anderson: Science {\bf 235} (1987) 1196.

\bibitem{Ogata} M. Ogata and H. Fukuyama: 
Rep. Prog. Phys. {\bf 71}  (2008) 036501. 

\bibitem{Lee} P. A. Lee, N. Nagaosa, and X.-G. Wen:  
Rev. Mod. Phys. {\bf 78} (2006) 17. 

\bibitem{Kotliar} G. Kotliar and J. Liu: Phys. Rev. B{\bf 38} (1988) 5142. 

\bibitem{Suzumura} Y. Suzumura, Y. Hasegawa, and H. Fukuyama: 
J. Phys. Soc. Jpn. {\bf 57} (1988) 2768. 

\bibitem{Zou} Z. Zou and P. W. Anderson: Phys. Rev. B{\bf 37} (1988) 627. 

\bibitem{Nagaosa} N. Nagaosa and P. A. Lee: Phys. Rev. Lett. {\bf 64} (1990) 2450. 

\bibitem{KK1} K. Kuboki, M. Yoneya, and H. Yamase: 
Physica C{\bf 470} (2010) S163. 

\bibitem{Yamase2} H. Yamase, M. Yoneya, and K. Kuboki: 
Phys. Rev. B{\bf 84} (2011) 014508. 

\bibitem{Inaba} M. Inaba, H. Matsukawa, M. Saitoh, and H. Fukuyama: 
Physica C {\bf 257} (1996) 299. 

\bibitem{Yamase} H. Yamase and H. Kohno: Phys. Rev. B{\bf 69} (2004) 104526. 

\bibitem{Himeda} A. Himeda and M. Ogata: Phys. Rev. B{\bf 60} (1999) R9935. 

\bibitem{Ren} Y. Ren, J-H. Xu, and C. S. Ting: 
Phys. Rev.  Lett. {\bf 74} (1995) 3680. 

\bibitem{Feder} D. L. Feder and C. Kallin:  Phys. Rev. B{\bf 55} (1997) 559. 

\bibitem{KKYano} K. Kuboki and K. Yano: 
J. Phys. Soc. Jpn. {\bf 81}  (2012) 064711. 

\bibitem{Dahl} E. K. Dahl and A. Sudb\o:  Phys. Rev. B{\bf 75}  (2007) 144504. 

\bibitem{KKSig} K. Kuboki adn M. Sigrist:
J. Phys. Soc. Jpn. {\bf 67} (1998) 2873.

\bibitem{Gorkov} L. P. Gor'kov:  
Sov. Phys. JETP {\bf 9} (1959) 1364. 

\bibitem{LeeNag} P. A. Lee and N. Nagaosa: Phys. Rev. B{\bf 46} (1992) 5621. 

\bibitem{DKim} D. H. Kim and P. A. Lee: Ann. Phys. {\bf 272} (1999) 130. 

\bibitem{Fenton} G. C. Psaltakis and E. W. Fenton:  J. Phys. C{\bf 16} (1983) 3913. 

\bibitem{Mura1} M. Murakami and H. Fukuyama: J. Phys. Soc. Jpn. {\bf 67}
 (1998) 2784. 

\bibitem{Mura2} M. Murakami: J. Phys. Soc. Jpn. {\bf 69}  (2000) 1113.  

\bibitem{Kyung} B. Kyung: Phys. Rev. B{\bf 62}  (2000) 9083. 

\bibitem {Apens} A. Aperis, G. Varelogiannis, P. B. Littlewood, and B. D. Simons:  
J. Phys.: Condens. Matter {\bf 20} (2008) 434235.  

\bibitem{KK2} K. Kuboki: J. Phys. Soc. Jpn. {\bf 68}  (1999) 3150. 

\bibitem{Berg2} F. S. Bergeret, A. F. Volkov, and K. B. Efetov: Phys. Rev. Lett. 
{\bf 86} (2001) 4096. 

\bibitem{Esch} M. Eschrig, J. Kopu, J. C. Cuevas, and G. Sch\"on: 
 Phys. Rev. Lett. {\bf 90}  (2003) 137003. 

\bibitem{Buz} A. I. Buzdin: Rev. Mod. Phys. {\bf 77} (2005) 935.

\bibitem{Berg} F.S. Bergeret,  A. F. Volkov, and K. B. Efetov:  
 Rev. Mod. Phys. {\bf 77}  (2005) 1321.
 

\end{thebibliography}
\end{document}